      [1999/12/01 v1.4c Il Nuovo Cimento]
\documentclass[arxiv,nocopyright]{cimento}

\def\babar{\mbox{\slshape B\kern-0.1em{\smaller A}\kern-0.1em 
  B\kern-0.1em{\smaller A\kern-0.2em R}}}

\title{Lessons from charm dynamics}
\author{Ayan Paul}
\instlist{\inst INFN, Sezione di Roma, I-00185 Roma, Italy}
\begin{document}

\maketitle

\begin{abstract}
We shall proceed to discuss the significance of charm dynamics in an era when the Standard Model is well established in its degrees of freedom but little understood in quite a few of its different parts. The study of decays and asymmetries in the decays of mesons with charm quantum numbers provide us possible insights into the yet uncharted dynamics of the Standard Model at the GeV scale. We will discuss some interesting channels with hadrons and/or leptons in the final state, the shortcomings of theoretical structures currently advocated and future experimental prospects.
\end{abstract}

\section{Introduction}

After more than a century of rigorous experimentations, all the ``fundamental''  degrees of freedom of the Standard Model (SM) have been discovered with the final saga being the discovery of the Higgs Boson. The stage now lies open for dynamics beyond what we have grown to understand as Standard. Yet, our understanding of a significant part of SM dynamics lacks solid grounding and leads the lack of predictability in many arenas that have been somewhat neglected in the race to discover theoretically established resonances. Charm dynamics is one such important arena which stands on the threshold that can be, at best, called ``pseudo-perturbative'' dynamics. Not only is it capable of giving us important insights into the workings of the SM in an area which is, and has always been, gray, but also serves as a testing ground for innovations in theoretical technologies beyond what is well established. 

Testing for the presence of new dynamics (ND) is now the central theme of our study of Particle Physics or High Energy Physics. Dynamics beyond the SM is very well motivated by various arguments such as: \\\\
$\bullet$ addressing the challenges of scale separation between electroweak and Planck scale,\\
$\bullet$ finding the key to the observed matter anti-matter asymmetry, \\
$\bullet$ explaining the flavour hierarchy in the SM, \\
$\bullet$ explaining the presence of massive neutrinos and their oscillation\footnote{This can be argued to be a part of SM.}, \\
$\bullet$ contemplations on the existence and representations of dark matter and dark energy, \\
$\bullet$ various arguments revolving around the statement: ``this cannot be all'', \\\\
to name a few. However, it is becoming increasingly obvious that ND will appear not with great fanfare but with great subtlety. While we continue on our quest to discover new degrees of freedom, it is becoming more and more important to focus on precision measurements in the flavour sector as the scales of dynamics probed is well known to be ``loop-enhanced'' and hence several orders of magnitude larger than what direct detection techniques can probe.

Charm dynamics serves as an interesting testing ground for the existence of relatively large ND effects because the signatures left by SM dynamics are tiny due to a very effective GIM/CKM suppression and the lack of a large hierarchy in the down type quark masses. However, when it is necessary to study percentage effects of ND, one needs a good handle on SM predictions, something that is lacking in general in charm dynamics. Hence the study of charm dynamics not only allows us to probe ND but also gain a more concrete understanding of the SM at the GeV scale. This, to say the least, has been an unsurmountable challenge in the past few decades due to the breakdown\footnote{Here ``breakdown'' is not tantamount to ``inapplicable''.} of our work-horses, the perturbation technologies, at the GeV scale. 

In the next few sections we will discuss few decay channels and asymmetries in charm dynamics that serve the dual purpose of providing a field for developing new theoretical technologies for dealing with the pseudo-perturbative dynamics of the SM and the study of ND intervention.

\section{Radiative mode: \boldmath $D^0\to\gamma\gamma$}
\label{sec:Dgg}

The branching fraction of the channel $D^0\to\gamma\gamma$ is yet 
to be measured. The current best experimental bound from the 
$\babar$ Collaboration \cite{Lees:2011qz} and the SM prediction \cite{Fajfer:2001ad,Burdman:2001tf,Paul:2010pq}, respectively, are
\begin{eqnarray}
{\rm BR}_{\rm exp}(D^0\to\gamma\gamma) < 2.4\times 10^{-6} \, (\rm 90\% \, C.L.)\;,\;
{\rm BR}_{\rm SM}(D^0\to\gamma\gamma) \sim (1-3)\times 10^{-8}.
\end{eqnarray}
The primary contribution to this comes from long distance (LD) effects 
\cite{Fajfer:2001ad,Burdman:2001tf}. 
The SM short distance (SD) estimate is several orders of magnitude smaller 
and is estimated at ${\cal O}(10^{-11})- {\cal O}(10^{-12})$. 
Hence, it is expected that long distance effects will dominate here. 

The $c\to u\gamma$ vertex is dominated by SM QCD effects \cite{Greub:1996wn} which is quite difficult for ND to overcome with SD dynamics. Moreover, even if this is possible the LD contribution dominates over the $c\to u\gamma$ vertex contribution by a few order of magnitude. Hence, this is not a good testing ground for ND. The measurement of this branching fraction provides possible validations of models used for estimating LD dynamics \cite{Fajfer:2001ad,Burdman:2001tf} and is  also important as it gives us an indirect measurement of the long distance 
contribution to $D^0\to\mu^+\mu^-$.

BESIII is projecting a reach of $5\times10^{-8}$ within the next $4-5$ years 
\cite{Asner:2008nq}. 
Hadronic machines are not supposed to be able to reach any reasonable 
precision in this channel and any further development has to be left to super
flavour factories.

\section{Leptonic mode: \boldmath $D^0\to\mu^+\mu^-$}
\label{sec:Dmumu}

There can be 
large new physics intervention in $D^0\to\mu^+\mu^-$. Due to a 
very effective GIM suppression the SM short distance contribution is 
very tiny \cite{Burdman:2001tf,Paul:2010pq} while the dominant contribution comes from long distance processes, primarily 
from the two photon unitary contribution $D^0\to\gamma\gamma\to\mu^+\mu^-$
\cite{Burdman:2001tf},
\begin{eqnarray}
&&{\rm BR^{SD}_{SM}}(D^0\to\mu^+\mu^-) \approx 6\times10^{-19},\\
&&{\rm BR^{LD}_{SM}}(D^0\to\mu^+\mu^-) = 
2.7\times10^{-5}\times {\rm BR}(D^0\to\gamma\gamma)\simeq2.7-8
\times10^{-13}.
\end{eqnarray}
It should be noted here that this contribution comes from the {\em total} 
branching fraction ${\rm BR}(D^0\to\gamma\gamma)$ {\em regardless} of whether 
the latter is generated by SM or ND. Hence a good measurement of 
${\rm BR}(D^0\to\gamma\gamma)$ will give us a very good control over the 
SM LD contribution in $D^0\to\mu^+\mu^-$.

ND intervention can be 
large in this channel \cite{Burdman:2001tf,Paul:2010pq,Paul:2012ab}. It is also possible to see ND contribution in this channel even if the 
analogous channels in the beauty system $B_{s(d)}\to\mu^+\mu^-$ yield 
measurements which are indiscernible from SM predictions \cite{Paul:2012ab}.

LHCb now sets a limit of $6.2(7.6)\times 10^{-9}$ with $90\%(95\%)$ confidence level 
with 0.9fb$^{-1}$ \cite{Aaij:2013cza}. The LHCb reach can be estimated 
to be another two or three orders of magnitude in the future with similar reach at ATLAS and CMS. This 
channel is measurable at super flavour factories only in the case of orders of magnitude enhancements from ND.

\section{Leptonic modes: \boldmath $D\to X_u l^+l^-$}
\label{sec:DXll}

The inclusive mode $D\to X_u l^+l^-$ is dominated by resonance contributions from $\rho$, $\phi$ and $\omega$. The exact branching fraction 
itself depends on the particular inclusive mode but is predicted to be about ${\cal O}(10^{-6}) - {\cal O}(10^{-7})$
\cite{Burdman:2001tf,Fajfer:2001sa}. At the charm scales the duality between quarks and hadrons breaks down significantly and hence predictions in exclusive modes no longer follow from those in inclusive modes. This has been highlighted by the recent LHCb upper bound \cite{Aaij:2013sua}
\begin{equation}
{\rm BR^{exp}}(D^+\to \pi^+ \mu^+\mu^-)< 7.3(8.3)\times 10^{-8}\;\;\;\textrm{at}\;\;90\%(95\%)\;\textrm{CL}.
\end{equation}
The dominant long distance contribution is estimated from $D^+\to\pi^+V:V\to \mu^+\mu^-$, where $V$ is $\phi,\rho$ or $\omega$. The short distance $c\to u l^+l^-$ contribution is three orders of magnitude 
smaller \cite{Burdman:2001tf,Fajfer:2001sa,Paul:2011ar}:
\begin{equation}
{\rm BR^{SD}_{SM}}(D\to X_u l^+l^-)\simeq 3\times 10^{-9}.
\end{equation}

The SM SD contribution is driven by the photon penguins which are no longer strictly local operators as they have  light quarks running in the loops. It is difficult 
for most generic ND to contribute to the branching fraction and enhance it 
by orders of magnitude \cite{Burdman:2001tf,Paul:2011ar}, except in very extreme parameter space of certain models \cite{Fajfer:2007dy} which are highly constrained from other flavour observables. 

What are significant in this mode for the intervention of ND are the asymmetries, namely, the 
forward-backward asymmetry $A^c_{\rm FB}$, the CP asymmetry $A^c_{\rm CP}$ 
and the CP asymmetry in the forward-backward asymmetry $A^{\rm CP}_{\rm FB}$. 
In the Standard Model these asymmetries are extremely tiny and are of 
${\cal O}(10^{-6})-{ \cal O}(10^{-4})$. 
These are relatively clean of long distance dynamics and even 
more so if the vector resonances are kinematically excluded from the 
invariant dilepton mass distribution. While generic ND can enhance both the $A^c_{\rm FB}$  
to ${\cal O}(1\%)-{\cal O}(10\%)$ and $A^c_{\rm CP}$ to ${\cal O}(1\%)$, 
they can enhance $A^{\rm CP}_{\rm FB}$ to ${\cal O}(10^3\%)-{\cal O}(10^4\%)$, 
i.e., well beyond the SM reach \cite{Paul:2011ar}. An interesting model independent analysis of possibly sizable CP asymmetry is presented in \cite{Fajfer:2012nr} driven by large chromomagnetic penguins, the same source which can drive a CP asymmetry in $D^0\to K^+K^-/\pi^+\pi^-$. This will also lead to a sizable CP violation in $D\to V\gamma$ \cite{Isidori:2012yx,Lyon:2012fk}
 
With current and projected statistics at the LHCb experiment, these modes 
might be within reach. These measurements are ideal for a flavour factory 
with its clean environment.

\section{Leptonic modes: \boldmath $D^0\to h_1h_2l^+l^-$}
\label{sec:Dhll}

While there are distinct experimental and theoretical disadvantages of studying four body modes, there is also the advantage of having more observables due to different kinematic combinatorics which can give us additional handles over not only SM dynamics but also the nature of ND through distributions and moments. 

The T-odd correlation leading to CP asymmetry in $D^0\to h^+h^-l^+l^-$ is driven by 
the asymmetry in $D^0\to h^+h^-$ $(h\rightarrow K,\pi)$. An enhanced asymmetry can be 
expected in $D^0\to h^+h^-l^+l^-$ for reasons analogous to the case of $K_L\to\pi^+\pi^-l^+l^-$ \cite{Sehgal:1992wm}. Estimates show that the asymmetry 
can be driven up to ${\cal O}(1\%)$ \cite{Bigi:2011em} while a more recent work shows that it 
can be as high as ${\cal O}(5\%)$\cite{Cappiello:2012vg}. 

Other modes that can be studied for such T-odd correlations are singly Cabbibo suppressed modes with $h^+=K^+, h^-=\pi^-$. However, the asymmetries here can be expected to be smaller as they are driven by indirect CP violation generated by  oscillations. The branching fraction of all these four body modes are around $10^{-6}$ or less \cite{Cappiello:2012vg} and would require 
quite a bit of statistics to probe into the asymmetries. A super flavour factory would have enough 
statistics to access this. LHCb can possibly explore the final states with a pair of charged hadrons and a pair of muons.

\section{Hadronic modes: \boldmath $D\to h_1h_2h_3h_4$}
\label{sec:Dhhhh}

The invariant mass distributions of different pairs of final state hadrons give a lot of observables for the study of SM and ND. The important kind of observables to study here are the T-odd correlations that are connected to CP violation as in $D\to h_1h_2l^+l^-$. The difference here is that in the latter channel final state interactions (FSI) can be ignored to corrections of ${\cal O}(\alpha)$. However, FSI becomes important in the fully hadronic four body final state. While T odd correlations can be generated even without CP violation, the latter would imply

\begin{eqnarray}
\langle \vec h_1 \cdot (\vec h_2 \times \vec h_3 ) \rangle _{\rm D} \neq    
-\langle \vec h_1 \cdot (\vec h_2 \times \vec h_3 ) \rangle _{\rm \bar D} \;,\;
\langle \vec h_1 \cdot (\vec h_2 \times \vec h_4 ) \rangle _{\rm D} \neq    
-\langle \vec h_1 \cdot (\vec h_2 \times \vec h_4 ) \rangle _{\rm \bar D} 
\end{eqnarray}
which can be extended to any other triple moment made out of a combination of $h_1,h_2,h_3$ and $h_4$ which are the kinematic directions of the corresponding hadron. Once a certain set of triple product has been chosen as a test of CP violation, {\em distribution} of another set can be used to probe the underlying dynamics.

It is also possible to study the differential decay rate with respect to the angle $\Phi$ between the $K^+ - K^-$ plane and the $\pi^+ - \pi^-$ plane in $D^0\to K^+K^-\pi^+\pi^-$ decay. For the meson and the anti-meson we have:
\begin{eqnarray}
\\
\nonumber\frac{d}{d\Phi} \Gamma (D^0 (\bar{D}^0) \to K^+K^- \pi^+\pi^-) &=& 
\Gamma _1(\bar{\Gamma}_1) \cos^2 \Phi + \Gamma _2 (\bar{\Gamma}_2) \sin^2 \Phi + 
\Gamma _3 (-\bar{\Gamma}_3)  \cos\Phi\sin\Phi.
\end{eqnarray}
Under T transformation, $\Gamma_3$ is odd as $\Phi\to-\Phi$. However, $\Gamma_3\neq0$ is not a clear signal for CP violation as FSI can produce that too. The unambiguous sign of CP violation is $\Gamma_3\neq\bar\Gamma_3$. The CP asymmetry can then be defined as
\begin{equation}
A_{\rm CP} = \frac{\int _0^{\pi /2}d\Phi \frac{d\Gamma}{d\Phi } 
-\int _{\pi /2}^{\pi}d\Phi \frac{d\Gamma}{d\Phi }  }{\int _0^{\pi}d\Phi \frac{d\Gamma}{d\Phi }}= 
\frac{2(\Gamma_3 - \bar{\Gamma}_3)}{\pi (\Gamma_1 + \Gamma_2 )}.
\end{equation}
Similarly, other angles can be defined like $\Phi$ between planes containing different pairs of the final state mesons from which more information can be gleaned. 

LHCb can well access $D^0\to K^+K^-\pi^+\pi^-$, $D^0\to \pi^+\pi^-\pi^+\pi^-$,  $D^0\to K^+\pi^-\pi^+\pi^-$ and their conjugate states. Recent results from LHCb  \cite{Aaij:2013swa} display consistency with CP conservation. However, in such analysis large statistics is of prime importance as distributions need to be studied. A charm factory would definitely have an edge here.

\acknowledgments
I would like to thank the organizers of IFAE 2013 for their hospitality and, along with the participants, for accommodating my inability to speak Italian. My participation in the conference was funded by the European Research Council under the European Union's Seventh Framework 
Programme (FP/2007-2013) / ERC Grant Agreement n.~279972.

\bibliography{IFAE_2013_AP.bib}{}
\bibliographystyle{varenna}

\end{document}